\documentclass{PoS}

\title{The Soft X-ray Imager on board {\it EXIST}}

\ShortTitle{The Soft X-ray Imager on board {\it EXIST}}

\author{\speaker{Lorenzo Natalucci}, Angela Bazzano, Francesca Panessa, 
Pietro Ubertini\\
        INAF-Istituto di Astrofisica Spaziale e Fisica Cosmica, Roma, Italy\\
        E-mail: \email{Lorenzo.Natalucci@iasf-roma.inaf.it}}

\author{Gianpiero Tagliaferri, Roberto Della Ceca, Gabriele Ghisellini, 
Giovanni Pareschi\\
         INAF-Osservatorio Astronomico di Brera, Italy}

\author{Gabriele Villa, Patrizia Caraveo, Mauro Fiorini, Michela Uslenghi\\ 
        INAF-Istituto di Astrofisica Spaziale e Fisica Cosmica, Milano, Italy}

\author{Jonathan E. Grindlay\\
       Harvard Smithsonian Center for Astrophysics, Cambridge, MA, USA}

\author{Brian Ramsey\\
NASA-Marshall Space Flight Center, Huntsville, AL, USA}

\abstract{
The Soft X-ray Imager (SXI) is one of the three instruments on board {\it EXIST}, a multi-wavelength
observatory in charge of performing a global survey of the sky in hard X-rays searching for Sup
er-massive Black Holes (Grindlay \& Natalucci, these Proceedings). One of the primary 
objectives of {\it EXIST} is also to study with unprecedented
sensitivity the most unknown high energy sources in the Universe, like high redshift GRBs,
which will be pointed promptly by the Spacecraft by autonomous trigger based on hard X-ray 
localization on board. The presence of a soft X-ray telescope with an effective area 
of $\sim950$~cm$^2$ in 
the energy band 0.2-3 keV and extended response up to 10 keV will allow to make broadband 
studies from 0.1 to 600 keV. In particular, investigations of the spectra components 
and states of AGNs
and monitoring of variability of sources, study of the prompt and afterglow emission 
of GRBs
since the early phases, which will help to constrain the emission models and finally, 
help the
identification of sources in the {\it EXIST} hard X-ray survey and the characterization 
of the transient events detected.
SXI will also perform surveys: a scanning survey with sky coverage
$\sim2\pi$ and
a limiting flux of $\sim5\times10^{-14}$~cgs plus other serendipitous.
}

\FullConference{The Extreme sky: Sampling the Universe above 10 keV - extremesky2009,\\
		October 13-17, 2009\\
		Otranto (Lecce) Italy}

\begin{document}

\section{Introduction}

%

\begin{figure}
  \begin{minipage}{0.5\textwidth}
    \resizebox{\hsize}{!}{\includegraphics{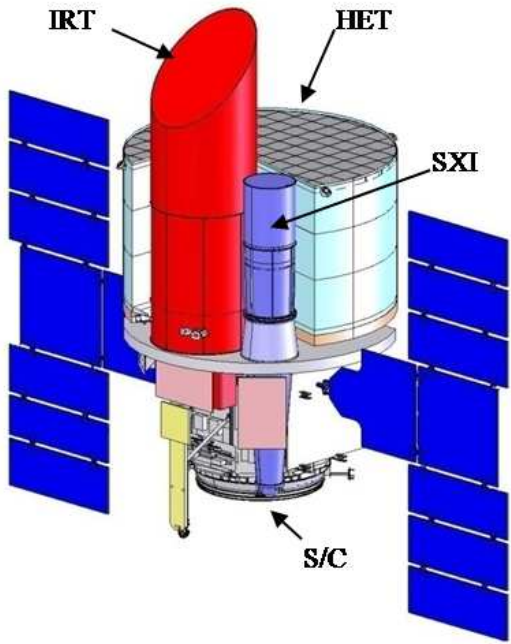}}
  \end{minipage}
  \begin{minipage}{0.45\textwidth}
    \caption{
The {\it EXIST} mission spacecraft with the three instrument complex: the High Energy telescope
(HET), a large field-of-view ($\sim90$deg) coded mask telescope and the 
Infrared Telescope (IRT) will be provided by the US whereas the Soft X-ray Imager (SXI)
will be contributed by Italy (ASI).
}
  \end{minipage}
\label{mission}

\end{figure}
%

Originally conceived as
a survey hard X-ray mission (Grindlay et al. 1995), {\it EXIST}   
has actually evolved into the concept of a 
multiwavelength mission covering the IR/optical and soft/hard X-ray bands
(see Grindlay \& Natalucci 2010). 
When fully funded, {\it EXIST} is scheduled for launch in mid 2017 with a EELV carrier. 
It will be put into a Low Earth Orbit (LEO) with lower inclination than Swift. 
In Fig. 1 is shown the baseline 
configuration of the {\it EXIST} mission.  
Its primary instrument is the High Energy
Telescope (HET), a wide field coded aperture instrument covering the 5-600 keV 
energy band and imaging sources
in a $70\times90$ deg$^2$ field of view with  
better than 20 arcsec 
positioning (Hong et al. 2010). The energy band
of HET overlaps with the soft X-ray range covered by  
SXI, 0.1-10 keV. The  
effective area of SXI is $\sim950$cm$^2$ at 1.5 keV and its focal length 
is 3.5m (Tagliaferri et al. 2009). 
At longer wavelenghts operates the IRT, an optical-IR aperture telescope
covering the 0.3-2.2 micron range with variable 
spectral resolution and high NIR sensitivity (AB=24 in 100s). The IRT can obtain 
spectra of GRB afterglows 
up to z$\sim20$ and make imaging and spectra of AGNs and transients 
discovered by the HET during the survey. 

%

\begin{figure}
  \begin{minipage}{0.5\textwidth}
    \resizebox{\hsize}{!}{\includegraphics{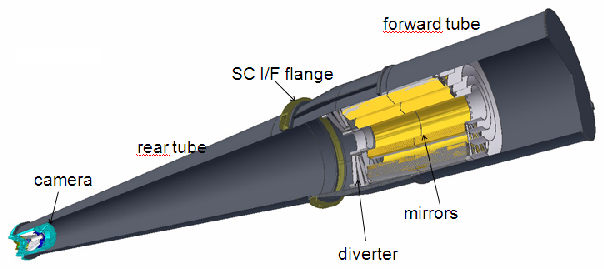}}
  \end{minipage}
  \begin{minipage}{0.45\textwidth}
    \resizebox{\hsize}{!}{\includegraphics{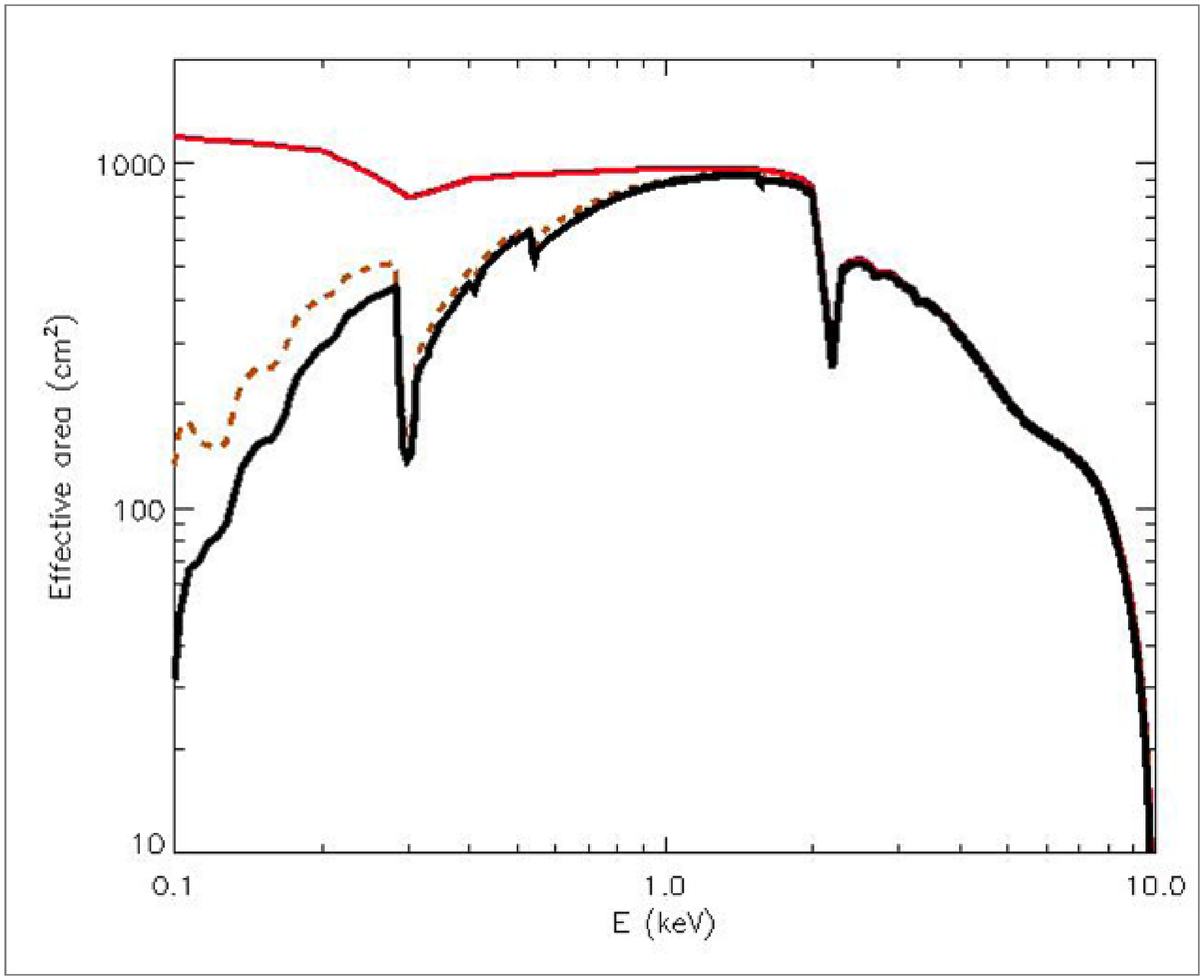}}
  \end{minipage}
\caption{
{\it Left}: view of the SXI instrument with details of parts (mirrors, camera and structure).
{\it Right}: the effective area of SXI (black). Also shown are the mirror effective area (red) and
its convolution with an XRT-like transmission filter (brown). 
}
\label{SXIwhole}

\end{figure}
%

\begin{table*}
\begin{center}
\caption{Design parameters of the SXI telescope}
\label{tableSXI}
\begin{tabular}
{   l   |  l   }             
\hline
Parameter      &   Baseline    \\ 
\hline
Mirror                    &    26 shells             \\
Angular Resolution        &    20~arcsec at 1 keV     \\
Energy Range              &    0.1-10 keV            \\
Diameter of Mirror        &    60cm    \\
Focal Lenght              &    3.5m    \\
Detector Type             &    {\em pn}-CCD  \\
Detector Size             &    3x3 cm$^2$   \\
FOV                       &    30x30 arcmin$^2$  \\
Energy Resolution         &    130eV at 6 keV   \\
Readout Speed             &    5-10 ms    \\
Effective Area            &    950 cm$^2$ at 1.5 keV, $>100$ cm$^2$ at 8 keV \\
Sensitivity ($10^{4}s$)    &    $2\times10^{-15}$ cgs   \\
\hline
\end{tabular}
\end{center}
\end{table*}

{\it EXIST} is a real multiwavelenght observatory for observations of GRBs and 
Supermassive Black Holes (SMXB)
as well as of many other types of transients and high energy sources. 
{\it EXIST} will take advantage of the same  
concept of operability of Swift with $\sim10$ times better
survey sensitivity in the high energy band from 0.1 to 600 keV. 
During the first two years of operation the {\it EXIST} observing time
will be mostly devoted to survey and GRB follow-up while the following 3 years 
are predicted to
be spent on pointed observations, taking full advantage of the presence of IRT and 
SXI. The first 
period survey will be performed by the HET pointing at the zenith 
with an offset of 30 degrees (towards 
the north and the south on alternate orbits, respectively) for an all-sky 
coverage each 3h.
This will allow detailed studies of obscured AGNs and to further study 
the accretion luminosity of SMBHs, as well as 
an "ultimate sensitivity" survey for Gamma-ray 
Bursts (Grindlay et al. 2009).

In this work we shall describe the Soft X-ray Imager, 
a substantially improved version of the Swift/XRT telescope, which has been 
added recently as part of the {\it EXIST} mission payload. 
The design of SXI in a "pre phase-A" style is in progress thanks to the provision of
an ASI grant following a competitive call for mission design studies. 

\section{Design of the SXI Instrument}

The proposed design of the SXI (see Fig. 2 and Table 1) is based on a Wolter type-I 
telescope consisting of a 
main mirror assembly with 26 nested
cells and a focal plane camera with CCD detector. The focal plane distance is 3.5m and 
the max.diameter of the mirrors is 60cm (giving 70cm on the telescope outer envelope). 
The telescope structure  is built around an I/F flange in titanium which is the 
interface to the satellite optical bench. The effective area of the instrument is shown in 
Figure 2 (right panel) and the main design parameters are listed in Table I.  

\subsection{The Mirrors} 

A detail of the mirror system is shown in Fig. 3 (left panel). The 
mirror shells are grouped in two
blocks in order to fulfill the desired effective area requirements. Possible space
for improving the effective area at medium energies is available but this must
consider weight constraints. 
A goal configuration with 
38 NiCo shells instead of 26 has also been studied, based on thicknesses as 
low as those designed for the shells of Simbol-X. We have then evaluated 
the on- and off-axis (10arcmin) effective 
area of the optics for the baseline and goal configurations. This is shown 
versus energy in Figure 3 
(right panel). The values of angular resolution on- and 
off-axis have been also computed using a conservative model for integration errors. 
The Half Power
Diameter (HPD) is estimated to be less than 20 arcsec throughout the whole 
field-of-view. 

%

\begin{figure*}[t!]
\resizebox{\hsize}{!}{\includegraphics[clip=true]{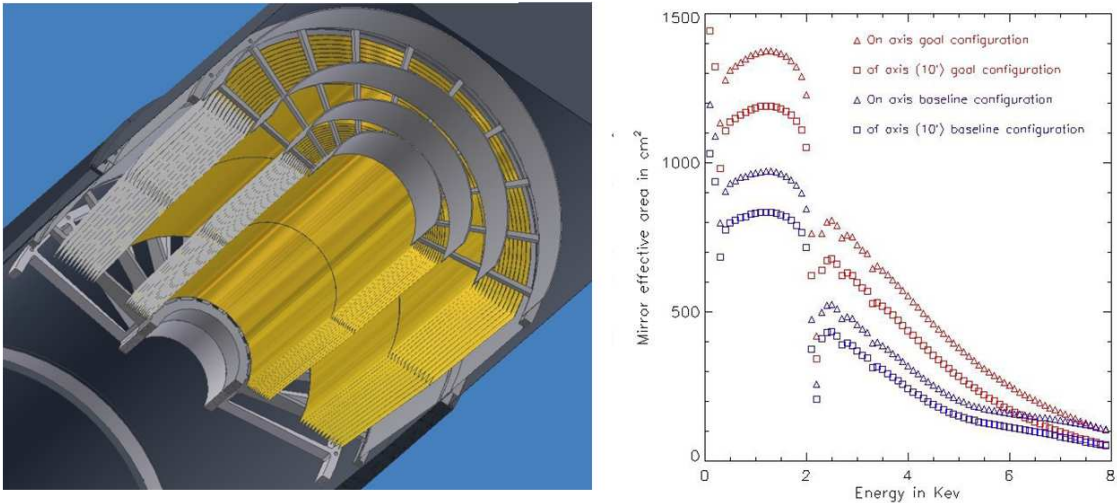}}
\caption{\footnotesize
{\it Left}: Detail of the baseline SXI mirrors with the 26 Ni nested shells. The on- and 
off-axis effective area are shown ({\it right}) for this configuration and an improved one
based on 36 NiCo shells.
}
\label{mirrors}

\end{figure*}
%

\subsection{The Detector and Camera}

The characteristics of the camera design (see Figure 4) are very similar to those of the
XRT and EPIC. 
In the current baseline the detector is a 3x3 cm$^2$ CCD sensor. Its  
Proximity Electronics is "suspended" within an Al shield and an active cooling system
will ensure the optimal temperature for CCD operation.  
In order to operate efficiently during the HET survey mode 
the sensor is required to have a frame
readout between 5 and 10 ms to compensate the relatively fast scanning. 
This can be achieved by the most recently developed CCDs as well as new generation
detectors like the Active Pixel Sensors (e.g., Str\"uder \& Meidinger 2009).
The camera comprises other mechanical subsystems 
like the Vacuum Chamber, Filter Wheel and Vacuum Door. In the present design the
Filter Wheel has 4 apertures, one of which is completely open, and one closed for CCD 
protection in case of high radiation. A third aperture consists of an X-ray filter 
to reduce the optical/UV contamination. Finally, 
the 4th  position consists of a closed 
position with a calibration source which illuminates directly the full detector 
in order to monitor 
its status, efficiency and the level of radiation damage.

%
\begin{figure}
  \begin{minipage}{0.6\textwidth}
    \resizebox{\hsize}{!}{\includegraphics{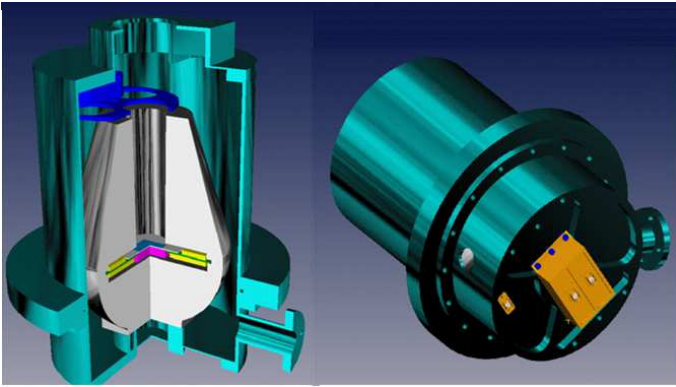}}
  \end{minipage}
  \begin{minipage}{0.35\textwidth}
    \caption{
3-D view of the inside of the SXI camera (left) and of the external part showing an
electronic box (right).
}
  \end{minipage}
\label{camera}

\end{figure}


\section{SXI scientific performances}

\subsection{Survey Sensitivity and Coverage}

During the zenith pointed scanning SXI will record X-ray events 
and the fast readout of its detector will allow to locate their direction.
The coverage of this survey will be half the sky and  
the resulting average exposure time for a source is of the
order of 200s/year, yielding a limiting sensitivity of 
$\sim5\times10^{-14}$~erg~cm$^{-2}$~s$^{-1}$
in two years. Since during the scanning the IRT does not operate, 
SXI will be able to improve the localization
of many faint HET sources from $\sim20$~arcsec to $\sim1-2$~arcsec.
On the other hand, during the second phase of the mission when {\it EXIST} 
is in pointing mode, 
SXI will help to obtain simultaneous coverage of the HET survey objects in the  
optical/NIR, soft X-ray to soft gamma-ray bands. In particular the SXI as well as the IRT will
measure source intensities and spectra for the hard X-ray sources that will be observed by HET down to a 
sensitivity limit of $\sim0.1$ mCrab in the hard X-ray band.

Serendipitous surveys by SXI during the 3 years of the inertial pointing phase 
will allow to cover $\sim1500$ square degrees down to the sensitivity
limit of $\sim2\times10^{-15}$~erg~cm$^{-2}$~s$^{-1}$.

\subsection{Study of Individual Objects}
SXI will be very useful for broadband studies as it will 
extend the high energy coverage of {\it EXIST} by more than one decade in    
energy. This is of outstanding importance for the study of accretion 
phenomena in particular to determine the state and physical parameters 
of a high number of AGNs and Galactic Black Holes. Furthermore SXI, 
together with the IRT
will help the optical identification and characterization  
of those sources detected by HET that are optically too faint to be
matched by existing optical catalogs (Della Ceca et al. 2010), like the highly
obscured, distant objects. In general SXI
will provide unvaluable data during the {\it EXIST} followup observations
of GRBs and during the inertial pointing observations performed in the second 
phase of the mission. These will include
SMBHs, transients of all types and GRBs. For GRBs, 
the detection of spectral features in the afterglow emission can be
used in the determination of the redshift, as well as to unveal the 
structure of the medium surrounding the central engine.     
 
\section{Conclusions}

The SXI telescope on board {\it EXIST} will take full advantage of the operational 
strategy adopted for the mission,
mostly based on surveys and fast follow-up of GRBs and transients. With SXI, {\it EXIST} 
is a real multiwavelength
observatory with a sensitive broadband coverage at high energies: 0.1-600 keV. 
SXI has an effective area of $\sim950$cm$^2$
at 1.5keV. It will perform wide area surveys (scanning and serendipitous) and 
sensitive observation of transient events
in the X-ray (e.g. GRB afterglows, AGN flares). 
It will also help the identification of HET
sources detected during the survey, the characterization of AGN states and 
the study of the absorbed 
(even Compton thick) AGN Universe.
The heritage of the Swift/XRT, XMM-Newton and INTEGRAL allows to conclude that the 
SXI performance
is appropriate to the profile of the {\it EXIST} mission as currently designed. 
Main improvements of the SXI with respect to Swift/XRT are: a 
factor $\sim10$ effective area and sensitivity, fast detector readout allowing 
full spectral imaging operation during the scanning survey.\\
 
{\bf Acknowledgements}
The Italian authors acknowledge the support of ASI by grant I/088/06/0.


\begin{thebibliography}{99}

\bibitem[1]{dellaceca} 
Della Ceca, R.  et al.\ 2010, 
"{\it The EXIST view of SuperMassive Black Holes in the Universe}, 
these Proceedings, accepted, arXiv:0912.3096 

\bibitem[2]{grindlay95} 
Grindlay, J.\ E.,  et al.\ 1995, 
{\it Energetic X-ray Imaging Survey Telescope},
Grindlay, J.\ E.,  et al.\ 1995, Proc. SPIE, Vol.2518, p.202

\bibitem[3]{grindlay09} 
Grindlay, J.\ E.,  et al.\ 2009, 
{\it GRB probes of the high-z Universe with EXIST},
AIPC, Vol.1133, p.18 (arXiv:0904.2210)

\bibitem[4]{grindlay10}
Grindlay, J.\ E. \& Natalucci, L. 2010,
{\it Surveying the Extreme Sky with EXIST},
these Proceedings, submitted 

\bibitem[5]{hong} Hong, J.~et al.\ 2010, 
{\it The High Energy Telescope on EXIST: hunting High Redshift GRBs and
other exotic transients}, these Proceedings, submitted

\bibitem[6]{struder}
Str\"uder, L. and Meidinger, N.\ 2009, {\it CCD Detectors}, in "The Universe in X-rays",
Springer Berlin Heidelberg, eds. J.E. Tr\"umper and G. Hasinger,
p.51 

\bibitem[7]{tagl} Tagliaferri, G.~et al.\ 2009,
{\it The Soft X-ray Imager (SXI) on board the EXIST mission}, 
Proc. SPIE, Vol.7437, p.743706 

\end{thebibliography}
\end{document}